\documentclass[aps,prb,10pt,twocolumn,superscriptaddress]{revtex4-1}

\usepackage{amsmath, amssymb}    					
\usepackage{graphicx}   							
\usepackage{xcolor}     							
\usepackage{hyperref}   							
\hypersetup{colorlinks=true,allcolors=blue}			
\usepackage{textcomp}								

\newcommand{\ddt}[0]{\frac{\partial}{\partial t}}
\renewcommand{\t}[1]{\textrm{#1}}
\newcommand{\nn}[0]{\nonumber\\}
\newcommand{\an}[0]{\allowdisplaybreaks\\}
\newcommand{\mbf}[1]{\mathbf{#1}}
\renewcommand{\k}[0]{\mathbf{k}}
\newcommand{\K}[0]{\mathbf{K}}

\newcommand{\R}[0]{\mathbf{R}}

\newcommand{\B}[0]{\mathbf{B}}
\newcommand{\q}[0]{\mathbf{q}}

\newcommand{\up}[0]{\uparrow}
\newcommand{\down}[0]{\downarrow}
\newcommand{\Jsd}[0]{J_{sd}}
\newcommand{\Jpd}[0]{J_{pd}}
\newcommand{\NMn}[0]{N_\t{Mn}}

\newcommand{\omMn}{\omega_{\t{Mn}}}
\newcommand{\ome}{\omega_{\t{e}}}

\newcommand{\gel}{g_\t{e}}

\newcommand{\gMn}{g_\t{Mn}}

\newcommand{\bs}[1]{\boldsymbol{#1}}

\newcommand{\me}{m_{\t{e}}}
\newcommand{\mh}{m_{\t{h}}}
\newcommand{\etae}{\eta_{\t{e}}}
\newcommand{\etah}{\eta_{\t{h}}}

\newcommand{\ud}{{\uparrow/\downarrow}}
\newcommand{\du}{{\downarrow/\uparrow}}

\newcommand{\omsf}[0]{\omega_\mathrm{sf}}
\newcommand{\ph}[1]{\phantom{#1}}

\begin{document}

\title{Phonon-induced quantum ratchet in the exciton spin dynamics in diluted magnetic semiconductors in a magnetic field}
\author{F. Ungar}
\affiliation{Theoretische Physik III, Universit\"at Bayreuth, 95440 Bayreuth, Germany}
\author{M. Cygorek}
\affiliation{Department of Physics, University of Ottawa, Ottawa, Ontario, Canada K1N 6N5}
\author{V. M. Axt}
\affiliation{Theoretische Physik III, Universit\"at Bayreuth, 95440 Bayreuth, Germany}

\begin{abstract}

Magnetically doped semiconductors are well known for their giant Zeeman splittings which can reach several meV even in relatively small external magnetic fields.
After preparing a nonequilibrium exciton distribution via optical excitation, the spin dynamics in diluted magnetic semiconductor quantum wells is typically governed by
spin-flip scattering processes due to the exciton-impurity exchange interaction.
Our theoretical calculations show that the giant Zeeman splitting in these materials in combination with the influence of longitudinal acoustic phonons lead to a quantum ratchet-type 
dynamics, resulting in an almost complete reversal of the carrier spin polarization at very low temperatures.
Furthermore, we find that the predictions of a much simpler rate-equation approach qualitatively agree with a more advanced and numerically demanding quantum 
kinetic description of the spin dynamics for a wide range of temperatures, although quantitative differences are noticeable.

\end{abstract}

\maketitle

\section{Introduction}
\label{sec:Introduction}

In quantum mechanics, a ratchet-type dynamics can ensue whenever there is a bias in favor of a specific scatting channel such that, e.g., a driving by an unbiased force
still causes a net current in the system \cite{Reimann_Quantum-Ratchets_1997, Ang_Quantum-ratchet_2015}.
An essential ingredient to the ratchet behavior is the breaking of time reversal symmetry \cite{Magnasco_Forced-thermal_1993}, which in solid state physics can be easily achieved
by applying a magnetic field.
Quantum ratchets have been experimentally realized and harnessed in a variety of systems, such as in atomic condensates \cite{Salger_Directed-Transport_2009}, all-optical 
systems \cite{Zhang_Experimental-demonstration_2015}, or at the single-electron level using the spin degree of freedom \cite{Costache_Experimental-Spin_2010}.
The latter approach falls into the field of spintronics, which aims to augment traditional electronic devices based on the manipulation of charge currents by including
the carrier spin \cite{Zutic_Spintronics-Fundamentals_2004, Awschalom_Challenges-for_2007, Dietl_A-ten_2010, Ohno_A-window_2010, Awschalom_Quantum-Spintronics_2013, 
Joshi_Spintronics_2016}.

A promising material class for bridging the gap between state-of-the-art semiconductor technology and spintronic devices are diluted magnetic semiconductors (DMSs)
\cite{Furdyna_Diluted-magnetic_1988, Furdyna_Semiconductor-and_1988, Kossut_Introduction-to_2010, Dietl_Dilute-ferromagnetic_2014}.
DMSs are doped with a small number of impurity ions that possess large magnetic moments, such as manganese, and which introduce a strong spin-dependent carrier-impurity
exchange interaction.
In this article, we theoretically show how a quantum ratchet emerges in the exciton spin dynamics in DMS quantum wells in a finite magnetic field when, in addition to the 
exchange interaction, the carrier-phonon scattering is accounted for.

Focusing on the subclass of II-VI DMSs, where the doping does not cause excessive charges in the system due to the isoelectronic nature of the impurity ions, we consider
the paramagnetic limit so that magnetic impurities at different sites in the crystal lattice can be considered to be independent.
Furthermore, the exciton resonance represents a regime which is often chosen in experiments using optical excitation \cite{Crooker_Optical-spin_1997, Camilleri_Electron-and_2001,
Smits_Excitonic-enhancement_2004, Murayama_Spin-dynamics_2006, BenCheikh_Electron-spin_2013} due to its spectral isolation and rich physics.
Since experiments are typically performed at very low temperatures \cite{Camilleri_Electron-and_2001, Vladimirova_Dynamics-of_2008, BenCheikh_Electron-spin_2013}, phonon scattering
is often not included in theoretical models and the focus instead lies on the description of spin-flip processes due to the carrier-impurity exchange interaction 
\cite{Bastard_Spin-flip_1992, Semenov_Electron-spin_2003, DasSarma_Temperatur-dependent_2003, Cywinski_Ultrafast-demagnetization_2007, Jiang_Electron-spin_2009,
Maialle_Exciton-spin_1993, Tsitsishvili_Magnetic-field_2006}.
Here, we show that longitudinal acoustic (LA) phonons, despite their negligible influence on the spin dynamics on short time scales, in conjunction with the spin-flip scattering in a finite magnetic field in fact lead to a drastically different longtime spin polarization which can be described in terms of a quantum ratchet.
Specifically, we find that a spin-conserving scattering of excitons causes an almost complete reversal of the initial spin polarization created by the optical excitation at very low temperatures.

To describe the spin dynamics, a rate-equation model is used which captures the spin-dependent exciton-impurity exchange interaction as well as the scattering of excitons due to
LA phonons.
We also compare the results of this model with calculations based on a more advanced treatment of the exchange interaction on a quantum kinetic level \cite{Ungar_Quantum-kinetic_2017},
which has recently been extended to account for LA phonon scattering \cite{Ungar_Phonon-impact_2018}.
That such a treatment of the exchange interaction beyond the mean-field level can be required for an accurate description of the physics in DMSs is supported by the pronounced
correlation effects exhibited by the material \cite{Ohno_Making-Nonmagnetic_1998, DiMarco_Electron-correlations_2013, Ungar_Many-body_2018}.
In the present study, our simulations reveal that both models yield qualitatively similar results provided that the phonon scattering is included.
However, this agreement is lost as soon as the exciton-phonon interaction is disregarded in the rate-equation model.

\section{Theoretical model}
\label{sec:Theoretical-model}

In this section, we present the main interaction parts of the Hamiltonian used for the description of the exciton spin dynamics in DMSs.
We also provide the resulting equations of motion in the Markov limit and discuss the energies involved in the spin-flip scattering due to the exciton-impurity exchange interaction.

\subsection{Interaction Hamiltonian}
\label{subsec:Interaction-Hamiltonian}

In DMSs, the dominant spin-flip mechanism is given by the $sd$ ($pd$) exchange interaction between $s$-type electrons ($p$-type holes) and the localized $d$-shell electrons 
of the magnetic impurities \cite{Kossut_Introduction-to_2010}.
It is described by \cite{Ungar_Quantum-kinetic_2017}
\begin{align}
\label{eq:H_m}
H_\t{m} =&\; \frac{\Jsd}{V} \sum_{\substack{I n n' \\ l l' \k \k'}} \mbf S_{n n'} \cdot \mbf s^\t{e}_{l l'} c^\dagger_{l \k} c_{l' \k'} e^{i(\k' - \k)\cdot\R_I} \! \hat{P}^{I}_{n n'}
	\nn
	&+ \frac{\Jpd}{V} \sum_{\substack{I n n' \\ v v' \k \k'}} \mbf S_{n n'} \cdot \mbf s^\t{h}_{v v'} d^\dagger_{v \k} d_{v' \k'} e^{i(\k' - \k)\cdot\R_I} \! \hat{P}^{I}_{n n'}
\end{align}
with the respective coupling constants $\Jsd$ and $\Jpd$ in a semiconductor with volume $V$.
The operator $c_{l\k}^\dagger$ ($c_{l\k}$) creates (annihilates) an electron in the $l$th conduction band with wave vector $\k$.
Analogously, $d_{v\k}^\dagger$ ($d_{v\k}$) refers to the respective hole operator in the valence band $v$.
The vector of electron (hole) spin matrices is given by $\mbf s^\t{e}_{l l'} = \bs\sigma_{l l'}$ ($s^\t{h}_{v v'} = \mbf J_{v v'}$) with the vector of Pauli matrices
$\bs\sigma_{l l'}$ and the vector of angular momentum matrices $\mbf J_{v v'}$, where $v,v' \in \{-3/2, -1/2, 1/2, 3/2\}$.
The impurity spin is decomposed into the vector of spin matrices $\mbf S_{n n'}$ with $n,n' \in \{-5/2, -3/2, ..., 5/2\}$ and the operator 
$\hat{P}^{I}_{n n'} = |I,n\rangle\langle I,n'|$, where $|I,n\rangle$ is the $n$th spin state of the $I$th impurity atom and $\R_I$ denotes its position.
This representation is advantageous for distinguishing impurity operators evaluated at the same and at different lattice sites, which in turn helps us to identify
the predominant correlations \cite{Thurn_Quantum-kinetic_2012}.

Apart from the magnetic exchange interaction, we also take the scattering of carriers with LA phonons into account.
As long as only optically excited low-energy excitons are considered, optical phonons can be disregarded for not too high temperatures since their energies are too high for an
allowed phonon emission process.
Furthermore, LA phonon scattering dominates the linewidth of optical spectra below temperatures of about $80\,$K \cite{Rudin_Temperature-dependent_1990}.
The carrier-phonon interaction is given by
\begin{align}
\label{eq:H_c-ph}
H_\t{c-ph} =&\; \sum_{\q \k} \Big( \gamma_\q^\t{e} c_{\k+\q}^\dagger c_\k b_\q + {\gamma_\q^\t{e}}^* c_\k^\dagger c_{\k+\q} b_\q^\dagger
	\nn
	&+ \gamma_\q^\t{h} d_{\k+\q}^\dagger d_\k b_\q + {\gamma_\q^\t{h}}^* d_\k^\dagger d_{\k+\q} b_\q^\dagger \Big).
\end{align}
Here, the creation (annihilation) operator for phonons with energy $\hbar\omega_\q^\t{ph}$ is denoted by $b_\q^\dagger$ ($b_\q$), where $\q$ contains the phonon momentum as well as the branch number.
We limit the description to bulk phonon modes and only consider deformation potential coupling \cite{Ungar_Phonon-impact_2018}.
Then, the coupling constants are given by
\begin{align}
\gamma_{\q,q_z}^\t{e,h} &= \sqrt{\frac{q \hbar}{2\rho V v}} D_\t{e,h}
\end{align}
for a semiconductor with density $\rho$, longitudinal sound velocity $v$, and deformation potential constants $D_\t{e,h}$ for the conduction and the valence band, respectively.
For the phonon dispersion, a linear relation $\omega_\q^\t{ph} = vq$ is assumed due to the small exciton center-of-mass momenta.

Apart from these interactions, we include the carrier kinetic energies, the Coulomb interaction responsible for the exciton binding, the light-matter coupling in the dipole 
approximation, as well as Zeeman terms for the carriers and the impurities that arise in an external magnetic field.
Furthermore, we account for the local potential mismatch in the lattice due to the doping with impurities by adding a nonmagnetic scattering contribution to the Hamiltonian 
in a form similar to Eq.~\eqref{eq:H_m} but without the spin part \cite{Cygorek_Influence-of_2017}.
Although this nonmagnetic impurity scattering does not affect the spin dynamics in a rate-equation model, it causes an enhancement of the correlation energy which can be captured 
by a quantum kinetic approach \cite{Ungar_Quantum-kinetic_2017}.
For explicit expressions of these Hamiltonian contributions, we refer the reader to Ref.~\onlinecite{Ungar_Trend-reversal_2018}.

\subsection{Rate-equation model}
\label{subsec:Rate-equation-model}

In the following, we restrict our considerations to systems with a sufficiently large energy splitting between heavy and light holes such that, using an optical excitation 
with $\sigma^-$ polarization resonant with the $1s$ heavy-hole exciton, the electron-spin part plays the dominant role in the exciton spin dynamics and the hole-spin part 
remains pinned along the growth direction \cite{Uenoyama_Hole-relaxation_1990, Bastard_Spin-flip_1992, Crooker_Optical-spin_1997}.
Thus, one can effectively assign two spin orientations to the excitons which follow directly from the electron spin.

If the $z$ axis is chosen to coincide with the growth direction of the quantum well as well as the direction of the applied magnetic field $\B$, the mean-field precession 
frequencies of electrons and Mn impurities are given by \cite{Ungar_Quantum-kinetic_2017}
\begin{subequations}
\label{eq:omega}
\begin{align}
\label{eq:ome}
\bs\ome =&\; \frac{1}{\hbar} \gel \mu_B \mbf B + \frac{\Jsd \NMn \langle S^z\rangle}{\hbar V} \mbf e_z,
	\an
\bs\omMn =&\; \frac{1}{\hbar} \gMn \mu_B \mbf B,
\end{align}
\end{subequations}
respectively.
In the above equations, $\gel$ ($\gMn$) is the electron (Mn) $g$ factor and $\mu_B$ denotes the Bohr magneton.
The Zeeman contribution due to the impurities scales with the number of Mn atoms given by $\NMn$ and depends on the $z$ component of the average impurity spin $\langle S^z\rangle$.
Finally, $\mbf e_z$ is the unit vector along the $z$ axis.
In this configuration, the $z$ components of these quantities yield the energetic splitting of the different exciton spin states which emerges as a result of the applied
magnetic field in combination with the successive alignment of the impurity spins.

To describe the spin dynamics, it is advantageous to use the spin-up ($n_{\omega_1}^\up$) and spin-down exciton density ($n_{\omega_1}^\down$) at a given frequency $\omega_1$.
From these variables, the $z$ component of the spin can be extracted via $s_{\omega_1}^z = \frac{1}{2} (n_{\omega_1}^\up - n_{\omega_1}^\down)$.
Since the focus of this paper lies on the Faraday configuration, where the magnetic field is aligned with the growth direction, in-plane spin components will remain zero
throughout the dynamics and can therefore be disregarded for typical optical excitation scenarios of spin-up excitons.
Treating all couplings in the Markov limit so that a Markovian theory (MT) is obtained, one then ends up with the following equations of motion \cite{Ungar_Quantum-kinetic_2017, Ungar_Trend-reversal_2018}:
\begin{widetext}
\begin{align}
\label{eq:Markov-equation}
\ddt n_{\omega_1}^\ud =&\; 
	\Gamma_{\omega_1} +  \frac{I \NMn M \Jsd^2}{2 \hbar^3 V d} \int_0^\infty \!\! d\omega \; \delta\big(\omega - (\omega_1 \pm \omsf)\big)
	F_{\etah 1s 1s}^{\etah \omega \omega_1} \big(b^\pm n_{\omega}^\du - b^\mp n_{\omega_1}^\ud\big)
	\nn
	&+ \int_0^\infty D(\omega) \Lambda_{1s 1s}^{\omega_1 \omega} \bigg[ \Theta\big(\omega - \omega_1 - \omega_{\omega-\omega_1}^\t{ph}\big)
	\Big( n_{\omega}^\ud \big(1+n^\t{ph}(\omega - \omega_1)\big) - n_{\omega_1}^\ud n^\t{ph}(\omega - \omega_1) \Big)
	\nn
	&+ \Theta\big(\omega_1 - \omega - \omega_{\omega_1-\omega}^\t{ph}\big)
	\Big( n_{\omega}^\ud n^\t{ph}(\omega_1 - \omega) - n_{\omega_1}^\ud \big(1+n^\t{ph}(\omega_1 - \omega)\big) \Big)\bigg].
\end{align}
\end{widetext}
Here, the magnetic moments due to the Mn ions are subsumed in the constants $b^\pm = \frac{1}{2}\big(\langle \mbf S^2 - (S^z)^2 \rangle \pm \langle S^z \rangle\big)$.
The equations are formulated in frequency space with the constant density of states $D(\omega) = \frac{VM}{2\pi\hbar d}$ and $\omega = \frac{\hbar K^2}{2M}$.
Furthermore, the optical generation rate of excitons is given by
\begin{align}
\label{eq:Gamma}
\Gamma_{\omega_1}(t) &= \frac{1}{\hbar^2} E(t) E_0 |M_\ud|^2 \phi_{1s}^2 \int_{-\infty}^t d\tau e^{-\frac{\tau^2}{2\sigma^2}} \, \delta_{\omega_1,0}
\end{align}
with $\sigma = \frac{t_\t{FWHM}}{2\sqrt{2\log 2}}$, where $t_\t{FWHM}$ is the time at the full-width half-maximum (FWHM) of the pulse $E(t) = E_0 \exp(-\frac{t^2}{2\sigma^2})$.
The optical spin selection rules are contained in the dipole matrix element $|M_\ud|^2$ and $\phi_{1s} = R_{1s}(r=0)$ denotes the radial part of the $1s$ exciton wave function
evaluated at the origin.
Regarding the phonon influence, $n^\t{ph}$ denotes a thermal phonon occupation given by $1/(\exp{(\hbar\Delta\omega/k_BT)}-1)$ and $\Theta(x)$ is the Heaviside 
step function.
Explicit expressions of the exciton form factor $F_{\etah 1s 1s}^{\etah \omega \omega_1}$ and the phonon matrix element $\Lambda_{1s 1s}^{\omega_1 \omega}$ can be found
in the Appendix.

Finally, we have introduced the exciton spin-flip scattering shift
\begin{align}
\label{eq:omsf}
\hbar\omsf := \hbar (\ome^z - \omMn^z),
\end{align}
where $\ome^z$ and $\omMn^z$ are the $z$ components of the precession frequencies given by Eqs.~\eqref{eq:omega}.
This quantity is a measure for the energy released or required during a spin-flip process of an exciton while simultaneously accounting for the resulting tilt of some Mn spins in
accordance with total spin conservation.
Instead of considering a time-dependent impurity spin density matrix, we focus on the limit of low exciton densities so that the Mn spin can be effectively treated as a spin
bath with a fixed temperature \cite{Ungar_Quantum-kinetic_2017}.

\begin{figure}
\centering
\includegraphics{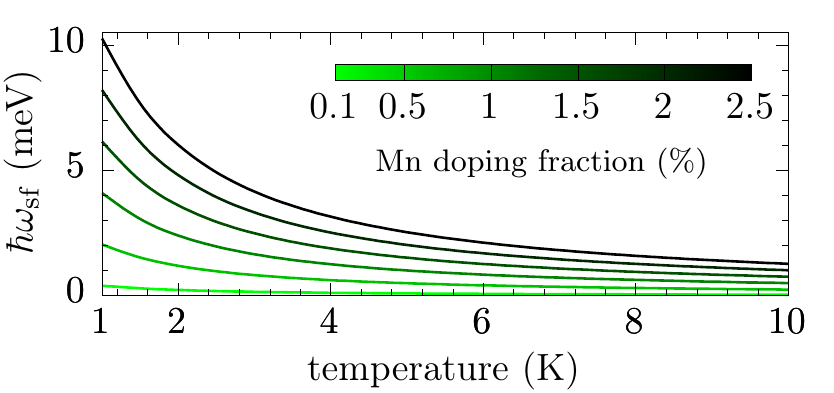}
\caption{Temperature dependence of the spin-flip scattering shift $\hbar\omsf$ for various Mn doping fractions of a $15\,$nm wide Zn$_{1-x}$Mn$_x$Se quantum well in an external 
		 magnetic field with a magnitude of $0.5\,$T.}
\label{fig:omegasf_vs_T}
\end{figure}

Typically encountered spin-flip scattering shifts are plotted in Fig.~\ref{fig:omegasf_vs_T} as a function of temperature for various impurity doping fractions.
The results are obtained for a $15\,$nm wide Zn$_{1-x}$Mn$_x$Se quantum well in an external magnetic field with a magnitude of $0.5\,$T.
If the temperature is low enough so that the majority of Mn spins are more or less aligned along the same direction, energies exceeding $10\,$meV can be reached provided the
doping fraction is high enough.
The fact that these energies by far exceed the standard Zeeman shifts typically encountered in solid state physics is known as the giant Zeeman effect 
\cite{Kossut_Introduction-to_2010, Furdyna_Diluted-magnetic_1988, Dietl_Dilute-ferromagnetic_2014}.
Since this effect relies on the overall magnitude of the average impurity spin [cf. Eq.~\eqref{eq:ome}], it strongly decreases with rising temperature since then the Mn spins
become more and more randomly oriented such that the average impurity spin approaches zero.

\section{Numerical results}
\label{sec:Numerical-results}

In the following, we present numerical calculations of the exciton spin dynamics in an external magnetic field with a focus on the influence of LA phonons.
All calculations are performed for a $15\,$nm wide Zn$_{0.985}$Mn$_{0.025}$Se quantum well in an external magnetic field with a magnitude of $0.5\,$T. 
The remaining material parameters are the same as in Ref.~\onlinecite{Ungar_Phonon-impact_2018}.
For the optical excitation, we model a Gaussian pulse with $100\,$fs FWHM resonant to the spin-up $1s$ exciton.
First, the impact of phonon scattering on the longtime spin dynamics in DMSs is studied using the rate-equation model.
Second, the results are compared with a more elaborate theoretical approach where the exciton-impurity interaction is treated beyond the Markov level.

\subsection{Phonon impact on the longtime behavior of the exciton spin}
\label{subsec:Phonon-impact-on-the-longtime-behavior-of-the-exciton-spin}

\begin{figure}
\centering
\includegraphics{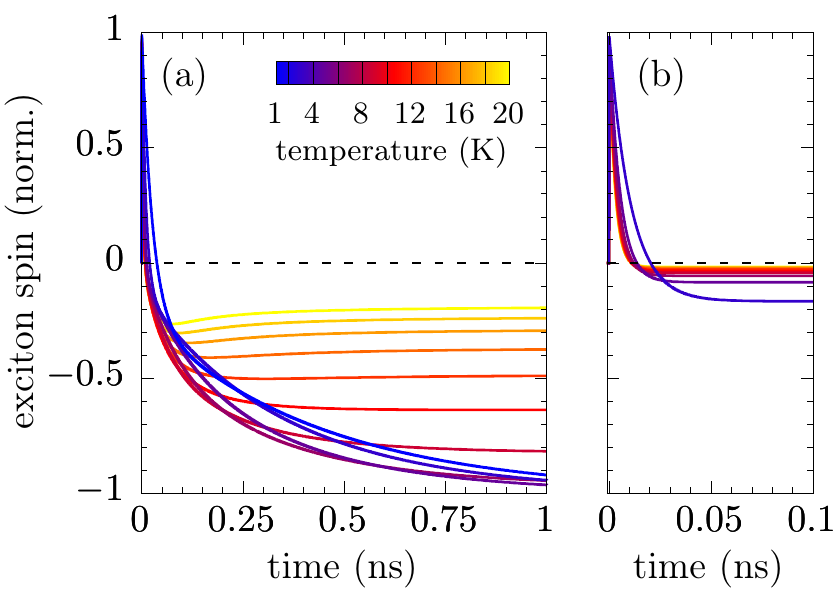}
\caption{Exciton spin dynamics after optical excitation for different temperatures (a) with LA-phonon scattering
		 and (b) without phonon scattering taken into account.
		 All curves are normalized with respect to the maximum spin polarization.}
\label{fig:T_var}
\end{figure}

Figure~\ref{fig:T_var}(a) shows the time evolution of the exciton spin for various temperatures under the influence of the magnetic exchange interaction as well
as phonon scattering, which are both treated on the Markov level corresponding to Eq.~\eqref{eq:Markov-equation}.
After the optical orientation of the exciton spins via a laser pulse which is resonant with the spin-up exciton ground state, a fast decay of the spin polarization occurs 
on a timescale of several picoseconds, followed by a slower decay until a stationary value is reached. 
Due to the external magnetic field in combination with the ensuing impurity spin polarization, the longtime value of the exciton spin is nonzero and instead shows a finite
spin-down polarization which, below $10\,$K, can be even higher than $50\%$ compared with the initial polarization.
Since the value of this polarization depends on the magnitude of $\langle S^z\rangle$, i.e., the degree of polarization of impurity spins \cite{Ungar_Trend-reversal_2018}, 
it also depends on the temperature used to calculate the impurity spin density matrix.
In the limit of either zero magnetic field or infinite temperature, the Mn spins are randomly oriented such that $\langle S^z\rangle = 0$ and the longtime value of the 
exciton spin is zero.
This explains the decrease of the absolute value of the longtime spin polarization with rising temperature observed in Fig.~\ref{fig:T_var}.

Interestingly, comparing Fig.~\ref{fig:T_var}(a) with Fig.~\ref{fig:T_var}(b) reveals that phonons completely alter the magnitude of the longtime spin polarization.
In Fig.~\ref{fig:T_var}(b), where the phonon scattering is not included in the calculations, one observes a maximal negative polarization of only $20\%$ for very low temperatures.
However, turning to Fig.~\ref{fig:T_var}(a), it becomes clear that this polarization increases to more than $90\%$ at very low temperatures when phonons are accounted for.
Furthermore, the overall influence of the temperature on the spin dynamics is strongly reduced for calculations without phonon scattering, as becomes evident by comparing
the wide spread of the longtime values in Fig.~\ref{fig:T_var}(a) with those observed in Fig.~\ref{fig:T_var}(b).
This last observation already hints at the mechanism behind the strong phonon impact on the spin dynamics: Since phonon absorption and emission processes are not equally
likely at low temperatures, an imbalance is created that is only overcome at elevated temperatures.
However, one has to keep in mind that phonons do not directly couple to the spin in our model since some kind of spin-orbit interaction is required to mix momentum and
spin scattering.

\begin{figure}
	\centering
	\includegraphics{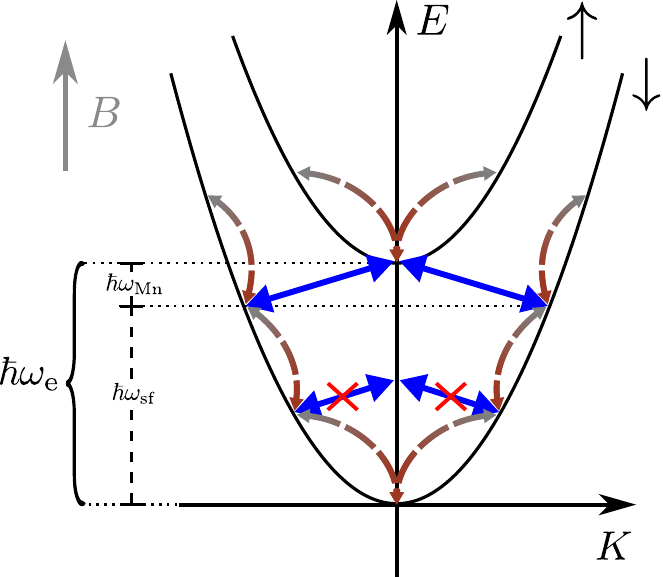}
	\caption{Sketch of the scattering processes in an external magnetic field $B$ between the spin-up ($\uparrow$) and spin-down ($\downarrow$) $1s$ exciton parabolas.
			 The spin-flip scattering due to the $sd$ exchange interaction is indicated by blue arrows while LA phonon scattering is sketched by arrows with a color gradient
			 from grey to brown, which in turn indicates the imbalance between phonon emission and absorption at low temperatures.
			 Forbidden spin-flip transitions are marked by red crosses and $\hbar\omsf$ denotes the spin-flip scattering shift given by $\hbar\ome - \hbar\omMn$, where
			 the Zeeman energy of excitons and impurities is denoted by $\hbar\ome$ and $\hbar\omMn$, respectively.}
	\label{fig:sketch}
\end{figure}

\begin{figure*}
\centering
\includegraphics{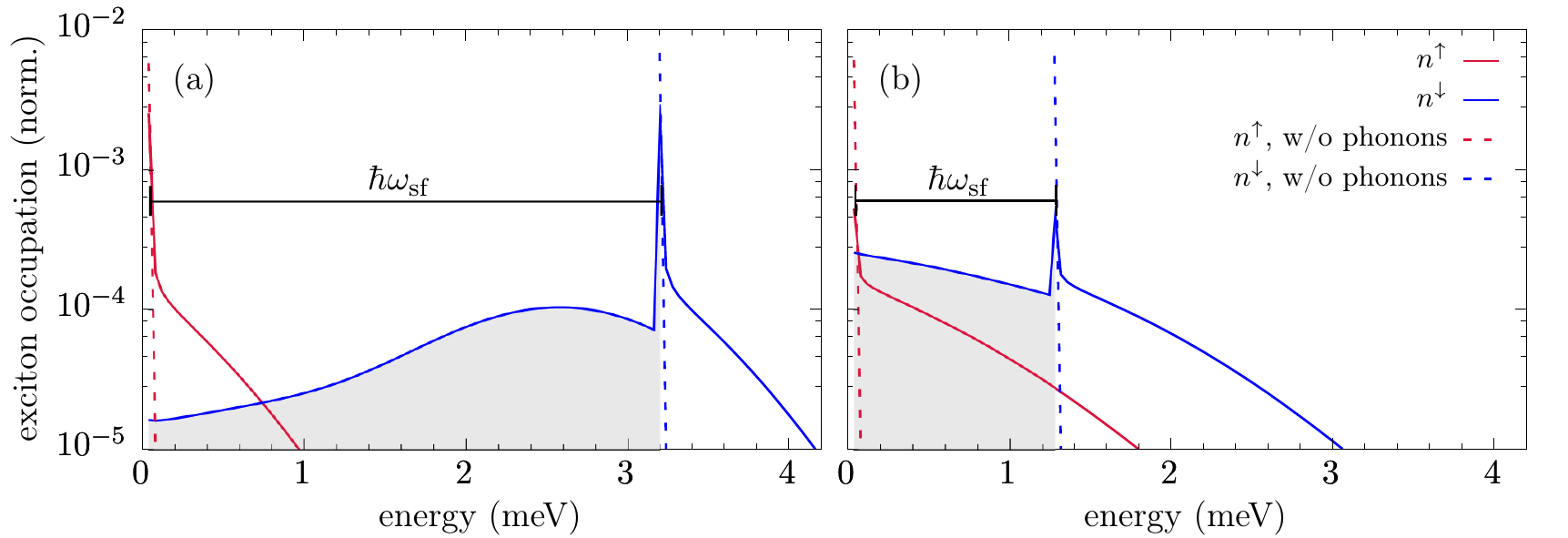}
\caption{Spin-up ($n^\up$) and spin-down ($n^\down$) occupation of the $1s$ exciton with and without scattering by LA phonons obtained $100\,$ps
		 after the optical excitation at a temperature of (a) $4\,$K and (b) $10\,$K, respectively.
		 For better visibility, the energy for each spin component is measured here from the bottom of the respective exciton parabola.
		 In fact, the minimum of the spin-up exciton parabola is shifted by $\hbar\ome^z$ (cf. Fig.~\ref{fig:sketch}).
		 Blocked spin-down states which cannot scatter back to the spin-up parabola are indicated by the grayed out area and $\hbar\omsf$ is
		 the spin-flip scattering shift induced by the external magnetic field including the giant Zeeman shift.
		 The occupation is normalized with respect to the exciton density after the pulse.}
\label{fig:ratchet_occup}
\end{figure*}

For a better understanding of the processes involved, the situation is sketched in Fig.~\ref{fig:sketch}, where two exemplary spin-flip transitions are indicated by blue arrows 
and the scattering with LA phonons is depicted by arrows with a color gradient, indicating the imbalance between phonon emission and absorption at the low temperatures studied here.
The spin-up parabola is Zeeman shifted by $\hbar\ome$ with respect to the spin-down parabola and the laser excitation is chosen to be resonant with this state, such that practically
only spin-up excitons are present shortly after the pump pulse.
After a spin-flip scattering to the spin-down parabola, the energy of an exciton decreases to $\hbar\omsf = \hbar\ome - \hbar\omMn$ in accordance with energy conservation, 
since an energy cost of $\hbar\omMn$ is required for the corresponding flop of an impurity spin (cf. also Sec.~\ref{subsec:Rate-equation-model}).
In the presence of phonons, excitons on the spin-down parabola can then further decrease their energy via phonon emission and effectively become captured in states with energies 
lower than $\hbar\omsf$ on the spin-down parabola.
However, a back scattering to the spin-up parabola is prohibited for such states since the final states would be below $\hbar\ome$, where the density of states is zero for spin-up
excitons. 
Such processes are exemplarily indicated by crossed-out blue arrows in Fig.~\ref{fig:sketch}.
Therefore, excitons can in this case only scatter back to the opposite spin state after absorbing a phonon with high enough energy, which is why, especially in the low-temperature 
limit where virtually no phonon absorption occurs, the back scattering to the spin-up state will be largely suppressed.

Indeed, looking at Fig.~\ref{fig:ratchet_occup} reveals that phonons have a decisive impact on the energetic occupation on the $1s$ exciton parabola.
Considering first the simpler situation without phonons, the first line of Eq.~\eqref{eq:Markov-equation} predicts a scattering of an initial occupation at $\hbar\omega_1$ to
states with energy $\hbar\omega_1 + \hbar\omsf$ and vice versa.
This means that, for optically generated excitons with nearly vanishing center-of-mass wave vector and, thus, negligible kinetic energy, a back-and-forth scattering between $E = 0$
and $E = \hbar\omsf$ is expected and no other states are involved.
In Fig.~\ref{fig:ratchet_occup}(a) and (b), where results for $T = 4\,$K and $T = 10\,$K are shown, respectively, the phonon-free results are indicated by dashed lines which 
show a separation of exactly $\hbar\omsf$.
The decrease of the spin-flip scattering shift observed in the figure is due to the increased temperature in accordance with Fig.~\ref{fig:omegasf_vs_T}.

Turning to the phonon influence, one has to be aware that phonon emission is prohibited for states at $E = 0$ since there are no exciton states with lower energy to scatter to.
Thus, phonon emission can only occur once a spin-flip scattering from the spin-up (red line) to the spin-down (blue line) state has taken place.
However, since a spin flip of an exciton also implies a change of its kinetic energy by the spin-flip scattering shift, phonons start to decrease the kinetic energy of excitons  via emission processes as soon as a spin flip occurs.
But because the energy of the exciton then inevitably falls below $\hbar\omsf$, there is suddenly no corresponding state with opposite spin available that can be reached via
exchange interaction scattering as its energy would have to be below $E = 0$.

All in all, the resulting dynamics behaves like a quantum ratchet, i.e., there is an imbalance between different scattering processes such that spin-down states are
strongly favored.
For the curves in Fig.~\ref{fig:ratchet_occup} where the phonon scattering is included, this means that the spin-down states with $E < \hbar\omsf$ are effectively blocked
and cannot scatter back to a spin-up state (cf. the grayed-out areas in Fig.~\ref{fig:ratchet_occup}).
This leads to a much larger spin-down occupation compared with the spin-up component already after $100\,$ps as depicted in the figure.
Note that, after $100\,$ps, the exciton system has not yet reached its thermal equilibrium for the parameters studied here.
In fact, a full thermalization of the exciton system due to the phonon influence only occurs on a nanosecond timescale.
The observation that the ratchet effect is apparently diminished with rising temperature can be explained by the combination of two effects.
First, the spin-flip scattering shift decreases with increasing temperature as shown in Fig.~\ref{fig:omegasf_vs_T}.
Second, phonon absorption becomes increasingly likely so that the scattering towards higher energies becomes possible.

It should be noted that the sizable spin-flip scattering shift due to the giant Zeeman effect in DMSs is an important ingredient for the quantum ratchet discussed here.
Since the typical Zeeman shifts in nonmagnetic semiconductors are rather small, the resulting energy gap between the spin-up and the spin-down parabola can be efficiently
bridged by phonons already at very low temperatures.
Thus, in DMSs, the large impact of phonons on the spin dynamics is a direct consequence of the magnitude of $\hbar\omsf$.

\subsection{Markovian vs. non-Markovian predictions}
\label{subsec:Markovian-vs-non-Markovian-predictions}

Up to this point, all scattering processes have been considered Markovian, i.e., on the single-particle level.
However, a single exciton not only interacts with one isolated Mn atom but rather with many impurities.
Previous theoretical works have demonstrated that the many-body nature of the $sp$-$d$ exchange interaction in fact manifests in non-Markovian features of the exciton 
spin dynamics \cite{Ungar_Quantum-kinetic_2017, Ungar_Trend-reversal_2018} as well as distinct features in optical spectra \cite{Ungar_Many-body_2018}, effects which
require a treatment of the exciton-impurity interaction beyond the single-particle level.

Concerning the quantum ratchet effect discussed in the previous section, the redistribution of exciton kinetic energies due to the scattering with LA phonons to values smaller 
than the spin-flip scattering shift is found to be of major importance.
This leads to the question whether the exciton-phonon interaction is the only one causing such a redistribution.
In fact, it turns out that a treatment of the carrier-impurity interaction beyond the mean-field level can also significantly change the energy of the carriers
beyond what is expected if the $sp$-$d$ exchange interaction is treated as a purely elastic process \cite{Ungar_Quantum-kinetic_2017, Ungar_Trend-reversal_2018, Thurn_Non-Markovian_2013}.
The reason why a change of the kinetic energy can occur is that a quantum kinetic treatment accounts for the energy-time uncertainty and, thus, does not strictly enforce energy
conservation on the single-particle level on short timescales.
Moreover, due to the many-body character of the exciton-impurity interaction, the effective single-particle exciton states are no longer the correct energy eigenstates of the 
system.
As it turns out, the latter effect is very important in DMS nanostructures and leads to the build-up of a negative many-body correlation energy that can reach values up to 
several meV per exciton \cite{Thurn_Non-Markovian_2013, Cygorek_Influence-of_2017, Ungar_Many-body_2018}.
Since this correlation energy persists even in the longtime limit, the carrier kinetic energy increases accordingly, leading to a pronounced broadening of the carrier distribution.

\begin{figure}
\centering
\includegraphics{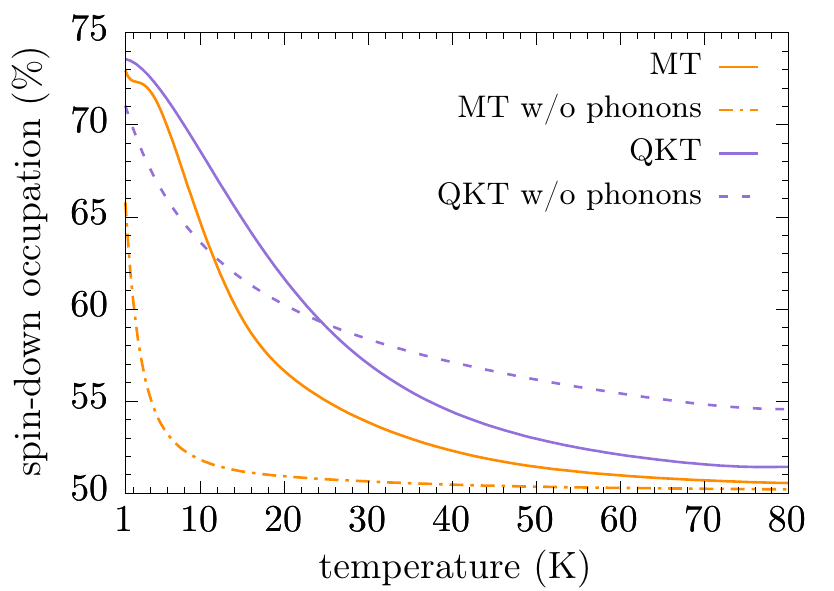}
\caption{Spin-down occupation after $100\,$ps with respect to the total exciton density calculated using the Markovian theory (MT) given by Eq.~\eqref{eq:Markov-equation} 
		 and a quantum kinetic theory (QKT), where the latter explicitly accounts for exciton-impurity correlations.
		 Also included are simulations where the scattering with phonons is switched off (w/o phonons).}
\label{fig:longtime_value_vs_T}
\end{figure}

To investigate the impact of many-body effects due to the $sp$-$d$ interaction on the ratchet-type dynamics discussed in the previous section, we look at the spin-down occupation
reached after $100\,$ps after the laser excitation as obtained by the purely Markovian theory described by Eq.~\eqref{eq:Markov-equation} and compare the results to those of a 
calculation based on a quantum-kinetic theory (QKT) of the exciton-impurity scattering, which has been recently developed in Ref.~\onlinecite{Ungar_Quantum-kinetic_2017} and 
explicitly takes correlations between excitons and impurities into account.
For a finite magnetic field and limiting the discussion to the exciton ground state, the necessary equations of motion for the exciton density, the spin, and the exciton-impurity
correlations can be found explicitly in Appendix A of Ref.~\onlinecite{Ungar_Trend-reversal_2018}.
In all cases, the exciton-phonon coupling is treated on the Markov level.

Figure~\ref{fig:longtime_value_vs_T} shows a comparison of the spin-down occupation at $100\,$ps normalized with respect to the total exciton density.
The results are obtained by the MT and the QKT for the case with and without phonons as a function of the temperature.
As has been already found in Fig.~\ref{fig:T_var}, phonons lead to large spin-down occupations especially at low temperatures when all interactions are treated on the Markov level.
For higher temperatures, the spin-down occupation quickly decreases and approaches the phonon-free calculation.
Note that a spin-down occupation of $50\%$ implies a balanced distribution of exciton spins in both spin channels, as would be expected without an external magnetic field.
Qualitatively, the QKT with phonon scattering shows a similar behavior compared with the MT, although the spin-down occupations as predicted by the QKT are consistently larger.

The influence of the many-body nature of the exciton-impurity interaction becomes most apparent when the phonon-free simulations are compared:
There, we find a significantly higher occupation of the spin-down states for the QKT compared with the MT for all considered temperatures.
The reason for this is that, while only a back-and-forth scattering between $E = 0$ and $E = \hbar\omsf$ can occur in the MT without phonons (cf. Fig.~\ref{fig:ratchet_occup}), 
the QKT captures a redistribution of excitons to other energies even without phonons in accordance with the build-up of a many-body correlation energy.
This means that, similar to the scattering due to phonons, there is now a finite possibility for spin-down excitons to occupy states below $E = \hbar\omsf$ such that they cannot 
directly scatter back to the spin-up state via the $sp$-$d$ exchange interaction and, once again, a quantum-ratchet effect occurs.
In fact, the scattering of excitons predicted by the QKT is rather significant for the impurity content considered here, which can be inferred from a comparison of the QKT results
with and without phonons.
Since the only change between the two results is the absence of phonon scattering in one case, the fact that the two curves lie close together compared to the respective results of
the MT means that the quantum kinetic scattering due to the exchange interaction provides the dominant contribution.
The reason why in the MT phonons seem to have a larger impact is the strict single-particle energy conservation which makes the ratchet effect impossible when only the spin-flip
exchange scattering on the Markov level is accounted for.

\section{Conclusion}
\label{sec:Conclusion}

We have investigated the phonon impact on the longtime value of the exciton spin polarization in DMS quantum wells after optical excitation
resonant with the exciton ground state.
The scattering of excitons to states with lower energies, which is the dominant phonon process at low temperatures, in combination with the much faster spin-flip scattering
due to the exciton-impurity interaction in DMSs leads to a ratchet-type spin dynamics that highly favors the energetically lower spin polarization.
For temperatures on the order of $5\,$K or below and a complete spin-up polarization after the excitation pulse, this effect is so strong that spin-down polarizations in excess 
of $90\%$ can be reached on a nanosecond timescale.

A rate-equation model, where all interactions are treated equally on the Markov level, allowed us to identify the quantum ratchet effect in terms of spin-flip
processes that become energetically blocked after phonon emission has occurred.
The fingerprint of the significant spin-down polarization is also visible in the exciton distribution as a function of energy.
A comparison of the spin-down occupation reached at $100\,$ps predicted by the MT with the results of a more advanced quantum kinetic description of the exciton
spin dynamics reveals an overall qualitative agreement, provided that the exciton-phonon scattering is included in both approaches.
However, quantitative differences are not negligible.

Without phonons, the quantum ratchet completely disappears in the MT while it is still present when the exciton-impurity scattering is treated on a quantum kinetic level.
The reason for this behavior is the redistribution of exciton kinetic energies due to a finite correlation energy, a many-body effect which is automatically included in the 
QKT but is not captured in the MT since it is beyond the single-particle level.
As this effect does not depend on the phonon scattering and is independent of the temperature, the quantum ratchet effect is retained by the QKT even in the absence of phonon
scattering.

\section{Acknowledgements}
\label{sec:Acknowledgements}

F.U. thanks A. Mielnik-Pyszczorski for helpful discussions and valuable input for designing Fig.~\ref{fig:sketch}.
Financial support of the Deutsche Forschungsgemeinschaft (DFG) through Grant No. AX17/10-1 is also gratefully acknowledged.

\section*{Appendix: Exciton and phonon form factors}
\label{app}

The form factor stemming from the projection onto the exciton basis reads \cite{Ungar_Quantum-kinetic_2017}
\begin{align}
\label{eq:angle-averaged-form-factors}
F_{\eta_1 1s 1s}^{\eta_2 \omega_1 \omega_2} =&\; 2\pi \! \int_0^{2\pi} \!\!\!\! d\psi \! \int_0^\infty \!\!\!\! dr \! \int_0^\infty \!\!\! dr' \, r r' R_{1s}^2(r) R_{1s}^2(r')
	\nn
	&\times J_0\big( \eta_1 K_{12}(\psi) r \big) J_0\big( \eta_2 K_{12}(\psi) r' \big),
\end{align}
where $K_{12} = |\K_1 - \K_2|$, $\psi$ denotes the angle between $\K_1$ and $\K_2$, and $K_i = \sqrt{\frac{2M\omega_i}{\hbar}}$.
In addition, $J_0(x)$ is the cylindrical Bessel function of order zero and $\eta_i = \frac{m_i}{M}$ with $i \in \{\t{e},\t{h}\}$ is the ratio between either the electron mass $\me$ 
or the heavy-hole mass $\mh$ and the exciton mass $M$, respectively.
Projected down to the energetically lowest quantum well confinement state, the coupling to the LA phonons can be subsumed in \cite{Ungar_Phonon-impact_2018}
\begin{align}
\Lambda_{1s 1s}^{\omega_1 \omega_2} =  
	\frac{P_{1s 1s}^{\omega_1 \omega_2} |\omega_1-\omega_2|}{\sqrt{1 - \frac{2Mv^2}{\hbar}}}
	\left|f\bigg(\frac{|\omega_1-\omega_2|}{v}\sqrt{1 - \frac{2Mv^2}{\hbar}}\bigg)\right|^2
\end{align}
with
\begin{align}
P_{1s 1s}^{\omega_1 \omega_2} =&\; \frac{2\pi}{\hbar \rho v^3 V} \bigg( D_\t{e}^2 F_{\etah 1s 1s}^{\etah \omega_1 \omega_2} + D_\t{h}^2 F_{\etae 1s 1s}^{\etae \omega_1 \omega_2} 
	\nn
	&+ 2 D_\t{e} D_\t{h} F_{-\etah 1s 1s}^{\ph{-}\etae \omega_1 \omega_2} \bigg).
\end{align}
We assume infinitely high quantum well barriers so that
\begin{align}
f(q_z) = \frac{\sin\big(\frac{q d}{2}\big)}{\frac{q d}{2}}\Big[1-\Big(\frac{q d}{2\pi}\Big)^2\Big]^{-1}.
\end{align}

\bibliography{references}
\end{document}